\documentclass{DISproc}

\usepackage{amsmath}
\usepackage{amssymb}
\begin{document}
\title{Jet-veto efficiencies at all orders in QCD}

\author{{\slshape Andrea Banfi$^1$}\\[1ex]
$^1$ University of Freiburg,
Hermann-Herder-Strasse 3, 79104 Freiburg, Germany}

\contribID{xy}

\doi  

\maketitle

\begin{abstract}  
  We present a next-to-leading logarithmic resummation for the
  jet-veto efficiency in Higgs production. We then discuss how this
  prediction affects the theoretical uncertainties in the region of
  transverse momenta of interest for Higgs searches at the LHC.
\end{abstract}

In searches of a Standard Model Higgs boson decaying into a pair of
$W$ bosons, it is customary to divide events with the requested
signature into bins corresponding to different jet
multiplicities~\cite{ATLAS:2011aa,Chatrchyan:2012ty}. In particular,
we concentrate on the 0-jet cross section, obtained by requiring that
there are no jets with transverse momentum $p_t$ larger than $p_{\rm
  t,veto}$. This 0-jet bin turns out to be less contaminated by $W$'s
originated from top-antitop production.

The 0-jet cross section can be computed at next-to-next-to-leading
order (NNLO) thanks to the fully differential calculations for Higgs
production of Refs.~\cite{Anastasiou:2005qj,Grazzini:2008tf}. The
question then arises on how to estimate the theoretical uncertainties
of the 0-jet cross section. It was already observed in
Ref.~\cite{Anastasiou:2008ik} that simultaneous variation of
renormalisation and factorisation scales leads to underestimating
theoretical uncertainties, which even vanish for $p_{\rm t,veto}
\simeq 25 {\rm GeV}$ . Therefore, the authors of that paper propose a
more sophisticated way to asses uncertainties based on those of the
inclusive cross section. The authors of Ref.~\cite{Stewart:2011cf} argued
further that the small scale uncertainty in the 0-jet cross section
was due to cancellations between two effects of different physical
origin. Indeed, $\sigma_{\rm 0-jets}(p_{\rm t,veto}) = \sigma_{\rm
  inclusive}- \sigma_{\rm \ge 1-jet}(p_{\rm t,veto})$. While the
inclusive cross section $\sigma_{\rm inclusive}$ is affected by a
large $K$-factor, the one for having more than one jet $\sigma_{\rm
  \ge 1-jet}(p_{\rm t,veto}) $ contains logarithmically enhanced
contributions $\alpha_s^n \ln^m (M_H/p_{\rm t,veto})$, with $m \le
2n$. Since these two effects are uncorrelated, they propose to
estimate the uncertainties on $\sigma_{\rm 0-jets}$ by just adding in
quadrature the uncertainties on $\sigma_{\rm inclusive}$ and
$\sigma_{\rm \ge 1-jet}$. We elaborate further on this ideas and write
the 0-jet cross section as the product of $\sigma_{\rm inclusive}$ and
the jet-veto efficiency $\epsilon(p_{\rm t,veto})$, defined as the
fraction of events such that all jets have a transverse momentum less
than $p_{\rm t,veto}$. We argue that the knowledge of higher and
higher orders for $\sigma_{\rm inclusive}$ does not help in reducing
the uncertainty in $\epsilon(p_{\rm t,veto})$, which rather reflects
our ignorance about logarithms $\alpha_s^n \ln^m (M_H/p_{\rm t,veto})$
of Sudakov origin, arising from a veto condition on real radiation. In
the following we then consider the uncertainties on $\sigma_{\rm
  inclusive}$ and $\epsilon(p_{\rm t,veto})$ as uncorrelated and we
concentrate on the efficiency only~\cite{Banfi:2012yh}.

At fixed order, the efficiency is defined in terms of the following
cross sections:
\begin{equation}
  \label{eq:1}
  \begin{split}
  &\sigma_{\rm inclusive} \equiv \sigma =
  \sigma_0+\sigma_1+\sigma_2+\dots\,,\\
  &\sigma_{\rm 0-jets}(p_{\rm t,veto}) = \Sigma(p_{\rm t,veto}) = 
  \sigma_0+\Sigma_1(p_{\rm t,veto})+\Sigma_2(p_{\rm t,veto})+\dots\,,
  \end{split}
\end{equation}
where $\sigma_i$ and $\Sigma_i(p_{\rm t,veto})$ are of relative order
$\alpha_s^i$ with respect to the Born cross section $\sigma_0$. It is
also useful to introduce the ``complementary'' cross sections $\bar
\Sigma_i(p_{\rm t,veto})$ as follows
\begin{equation}
  \label{eq:2}
  \bar \Sigma_i(p_{\rm t,veto}) = -\int_{p_{\rm t,veto}}^{\infty}
  \!\!\!\!\!\! d
  p_t\frac{d \Sigma_i(p_t)}{dp_t}\,,\qquad
  \Sigma_i(p_{\rm t,veto}) = \sigma_i+\bar \Sigma_i(p_{\rm t,veto}) \,.
\end{equation}
We remark that at the moment the perturbative expansion of these cross
section is known up to relative order $\alpha_s^2$.
We now identify three schemes that we believe cover the possibilities
to construct a jet-veto efficiency starting from the above cross
sections:
\begin{equation}
  \label{eq:3}
  \begin{split}
    \epsilon^{(a)}(p_{t,{\rm veto}}) &=
    \frac{\sigma_0+\Sigma_1(p_{t,{\rm veto}})+\Sigma_2(p_{t,{\rm
          veto}})}{\sigma_0+\sigma_1+\sigma_2}\,, \\
\epsilon^{(b)}(p_{t,{\rm veto}}) &= \frac{\sigma_0+\Sigma_1(p_{t,{\rm
      veto}})+\bar\Sigma_2(p_{t,{\rm veto}})}{\sigma_0+\sigma_1} \,,\\
\epsilon^{(c)}(p_{t,{\rm veto}}) &= 1+\left(1-\frac{\sigma_1}{\sigma_0} \right)\frac{\bar\Sigma_1(p_{t,{\rm veto}})}{\sigma_0}+\frac{\bar\Sigma_2(p_{t,{\rm veto}})}{\sigma_0}\,.
  \end{split}
\end{equation}
We observe that all these prescriptions differ only at order
$\alpha_s^3$. Each of them has its own meaning. Scheme~(a) is the
naive definition of the efficiency as the ratio between the 0-jet
cross section and the inclusive cross section. Scheme~(b) can be
motivated by the fact that the jet-veto efficiency can be seen as one
minus the probability of having one jet with $p_t>p_{\rm
  t,veto}$. This gives, at present accuracy, $\epsilon^{(b)}(p_{\rm
  t,veto}) = 1 - \sigma_{\rm \ge 1-jet}(p_{\rm t,veto})^{\rm NLO}
/\sigma_{inclusive}^{\rm NLO}$.  Finally, scheme~(c) corresponds to
the strict fixed order expansion of the efficiency.

In the following we use these three schemes as an extra handle,
besides renormalisation and factorisation scale variations, to
quantify the uncertainties on the jet-veto efficiency. Indeed, if we
compute $\epsilon(p_{\rm t,veto})$ for the three schemes, we obtain
very different predictions. Namely, including also independent
variation of renormalisation and factorisation scale in the range
$M_H/4 \le \mu_R, \mu_F \le M_H$ with $1/2 \le \mu_R/\mu_F \le 2$, in
the region of interest for experimental studies ($p_{\rm t,veto}=25
\mathrm{GeV}$ for ATLAS and $p_{\rm t,veto} = 30 \mathrm{GeV}$) for
CMS, the spread in fixed-order predictions for the jet-veto efficiency
is around 30\%~\cite{Banfi:2012yh,Dittmaier:2012vm}. This is not
observed in Z production, where all three schemes basically coincide.

Since part of this bad convergence can be attributed to the presence of
large logarithms of soft-collinear origin, it is useful to see how the
uncertainty changes when performing an all-order resummation of such
logarithms. At next-to-next-to-leading logarithmic (NLL) accuracy,
which amounts in controlling all terms $\alpha_s^n \ln^n(M_H/p_{\rm
  t,veto})$ in $\ln \epsilon(p_{\rm t,veto})$, this is possible with
the automated resummation program CAESAR~\cite{Banfi:2004yd}. In particular,
if jets are to be found everywhere in rapidity, CAESAR tells us that
the jet-veto efficiency is resummable within NLL accuracy, and has the
form
\begin{equation}
  \label{eq:4}
\epsilon(p_{t,{\rm veto}}) \sim {\cal L}_{gg}(p_{t,{\rm veto}}) \,
e^{-R(p_{t,{\rm veto}})} \,\mathcal{F}(R')\,,
\qquad
R' = - \,p_{t,{\rm veto}} \,\frac{d R(p_{t,{\rm veto}})}{d p_{t,{\rm veto}}}\,,
\end{equation}
where $ {\cal L}_{gg}(p_{t,{\rm veto}})$ is the gluon-gluon
luminosity, evaluated at the factorisation scale $p_{\rm t,veto}$, and
$R(p_{\rm t,veto})$ is the Sudakov exponent
\begin{equation}
  \label{eq:5}
  R(p_{t,{\rm veto}}) = 2 C_A \!\int_{p^2_{t,{\rm
        veto}}}^{M^2_H}\frac{dk_t^2}{k_t^2} \frac{\alpha^{\rm
      CMW}_s(k_t)}{\pi}\left[\ln\frac{M_H}{k_t}-\frac{4 \pi
      \beta_0}{C_A}\right]\,,\quad
\beta_0= \frac{11 C_A - 4 T_R n_f}{12 \pi}\,,
\end{equation}
where $\alpha^{\rm CMW}_s(k_t)$ is the physical coupling of
Ref.~\cite{Catani:1990rr}. While $R(p_{\rm t,veto})$ contains virtual
corrections only, the function $\mathcal{F}(R')$ accounts for multiple
soft-collinear real emissions. For a perfectly factorisable observable
$V(k_1,\dots, k_n) = \max_i\{V(k_i)\}$ we have
$\mathcal{F}(R')=1$. It turns out that, for small
$p_{\rm t,veto}$, if jets are defined with a $k_t^{2p}$-algorithm
(anti-$k_t$, Cambridge-Aachen, $k_t$), for emissions widely separated
in rapidity no recombination can occur. Therefore, it is only the
hardest gluon that contributes to the jet-veto efficiency, and
therefore $\mathcal{F}(R')=1$. A resummed prediction as the one in
Eq.~\eqref{eq:4} contains an extra source of theoretical
uncertainties. Indeed one can decide to resum $\ln(Q/p_{\rm t,veto})$
instead of $\ln(M_H/p_{\rm t,veto})$, where $Q$ is an arbitrary
``resummation'' scale we choose to vary in the range $M_H/4 \le Q \le
M_H$. Finally, to be able to present resummed predictions for the
jet-veto efficiency, we have to match the efficiency in
Eq.~\eqref{eq:4} with its expression at order $\alpha_s^2$. Therefore,
we introduce three matching schemes, defined in such a way that for
$p_{\rm t,veto} \ll M_H$ the matched efficiency reduces to the
expression in Eq.~(\ref{eq:4}), whilst for $p_{\rm t,veto} \sim M_H$,
it approaches its fixed order expression in any of the three schemes
introduced in Eq.~(\ref{eq:3}). The matching scheme gives us then an
extra handle to estimate theoretical uncertainties. 

\begin{figure}[h]
  \includegraphics[width=.5\textwidth]{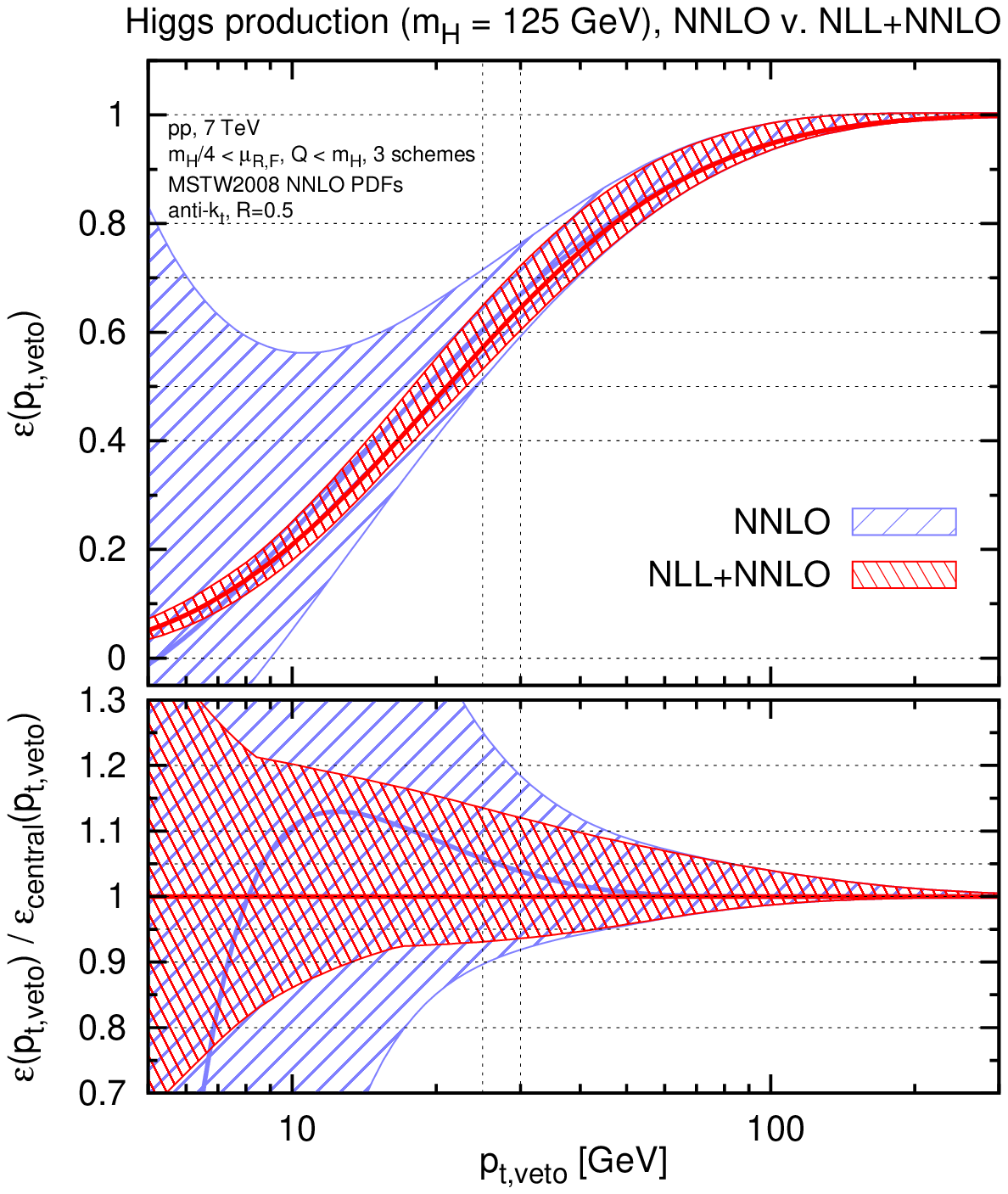}
  \includegraphics[width=.5\textwidth]{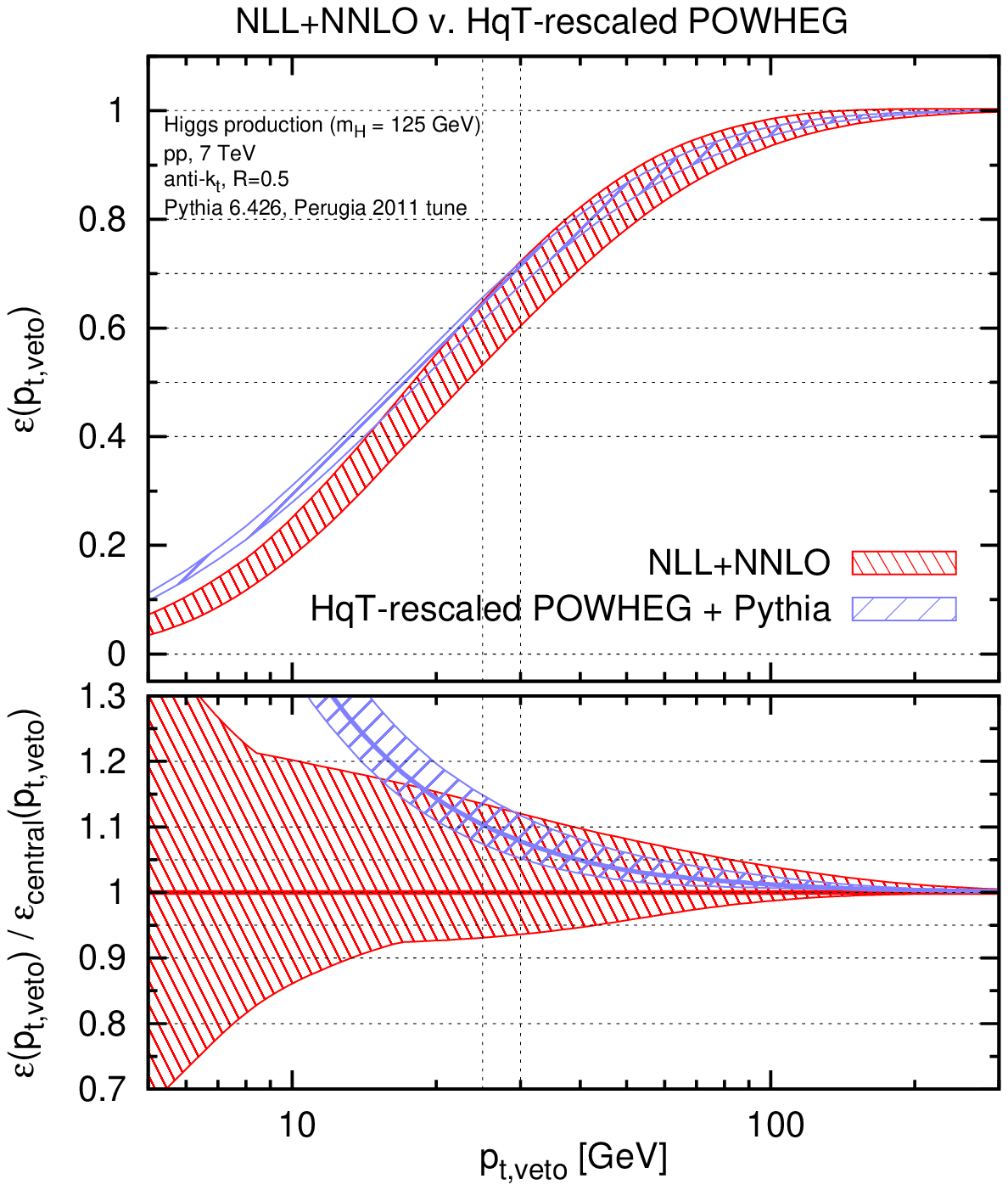}
  \caption{Left: the jet-veto efficiency for Higgs production at order
    $\alpha_s^2$ (NNLO) and matched to NLL resummation
    (NLL+NNLO). Right: NLL+NNLO efficiency compared to POWHEG rescaled
  using HqT.}
  \label{fig:eps-NLL}
\end{figure}
We now present results for the matched jet-veto efficiency, at the LHC
with $\sqrt s=7$ TeV, with jets clustered using the anti-$k_t$
algorithm with $R=0.5$, and for $M_H=125$ GeV. I order to estimate
theoretical uncertainties, we identify a ``central'' prediction, the
efficiency computed with matching scheme~(a) and with all scales
$Q,\mu_R,\mu_F$ equal to $M_H/2$. We then vary one scale at a time for
scheme~(a) in the range $[M_H/4,M_H]$, and vary the matching schemes
using $Q=\mu_R=\mu_F=M_H/2$. In this way we believe we do not double
count uncertainties. The predictions corresponding to this choice are
shown in Fig.~\ref{fig:eps-NLL}.  We observe that NLL resummation
helps reducing the uncertainties in the jet-veto efficency. Indeed,
for $p_{\rm t,veto}$ between 25 and 30 GeV, they move from 20\% (pure
NNLO) down to 10\% (NLL+NNLO). This improvement is not as huge, as is
for lower values of $p_{\rm t,veto}$, and reflects the fact that in
this intermediate region $\ln (M_H/p_{\rm t,veto})$ is not large
enough to guarantee that resummation effects dominate. We notice also
that here the uncertainty is dominated by the difference between
matching schemes. Since this difference is formally NNNLL, we do not
expect that a NNLL resummation could considerably help reducing the
theoretical uncertainty. Finally, we compare our predictions with the
Monte Carlo event generator that is currently used by CMS and ATLAS to
estimate the jet-veto efficiency. This is POWHEG~\cite{Alioli:2008tz}
interfaced to PYTHIA~\cite{Skands:2010ak}, rescaled in such a way that
it agrees with the Higgs $p_t$ spectrum computed at NNLL+NNLO accuracy
with the program HqT~\cite{Bozzi:2005wk}. The uncertainties in
POWHEG+PYTHIA are estimated by following the recommendation of
Refs.~\cite{Dittmaier:2012vm,Campbell:2012am}, i.e.~varying
renormalisation and factorisation scales independently around $M_H/2$
and fixing the parameter \texttt{hfact} to $h=M_H/1.2$. We observe
good agreement between our NLL+NNLO and POWHEG+PYTHIA in the region of
$p_{\rm t,veto}$ of interest. However, at lower values of $p_{\rm
  t,veto}$, we find that NLL+NNLO predictions tend to give lower
efficiency than that obtained with POWHEG+PYTHIA. We remark that the
same trend is observed when comparing NLL+NLLO to predictions obtained
with other Monte Carlo event generators.

To conclude, we have investigated how a NLL resummation for the
jet-veto efficiency affects the theoretical uncertainty on this
quantity. It would be very interesting to see how these findings
change after a NNLL resummation. In Ref.~\cite{Banfi:2012yh} we have
computed (for $R < \pi$) the part of NNLL resummation that depends on
the jet radius. The remaining NNLL contributions could be obtained by
relating the jet-veto efficiency to the Higgs $p_t$ spectrum (see for
instance~\cite{Becher:2012qa}). We hope to complete this study soon.

{\bf Note Added.} The NNLL resummation for the jet-veto efficiency in
Higgs and Drell-Yan production has been recently completed in
Ref.~\cite{Banfi:2012jm}.

\section*{Acknowledgements}
The work presented here has been done in collaboration with Gavin Salam and
Giulia Zanderighi. I am also grateful to Ian Winter for the invitation to the Workshop.


{\raggedright
\begin{footnotesize}



\end{footnotesize}
}


\end{document}